\documentclass[conference]{IEEEtran}

\usepackage{graphicx}
\usepackage{cite}
\usepackage{amsfonts}
\graphicspath{{eps/}} \DeclareGraphicsExtensions{.eps}

\usepackage[cmex10]{amsmath}
\usepackage{amssymb}
\usepackage{hyperref}
\usepackage{amsmath}
\usepackage{bm}
\usepackage{mathrsfs}
\usepackage{algorithm}
\usepackage{algorithmic}

\IEEEoverridecommandlockouts

\ifCLASSINFOpdf

\else

\fi

\begin{document}
%
\title{Energy Efficient Coordinated Beamforming for Multi-cell MISO Systems
\thanks{Corresponding author: Ling Qiu, lqiu@ustc.edu.cn.}}


\author{\IEEEauthorblockN{Yi Huang$^\dag$, Jie Xu$^\ddag$, and Ling Qiu$^\dag$}\\
\IEEEauthorblockA{
$^\dag$University of Science and Technology of China (USTC), Hefei, China, 230027\\
$^\ddag$Department of Electrical and Computer Engineering (ECE), National University
of Singapore (NUS), Singapore, 117576\\
Email: yihuang@mail.ustc.edu.cn, jiexu.ustc@gmail.com, lqiu@ustc.edu.cn}}


%


\maketitle

\begin{abstract}
In this paper, we investigate the optimal energy efficient coordinated beamforming in multi-cell multiple-input single-output (MISO) systems with $K$ multiple-antenna base stations (BS) and $K$ single-antenna mobile stations (MS), where each BS sends information to its own intended MS with cooperatively designed transmit beamforming. We assume single user detection at the MS by treating the interference as noise. By taking into account a realistic power model at the BS, we characterize the Pareto boundary of the achievable energy efficiency (EE) region of the $K$ links, where the EE of each link is defined as the achievable data rate at the MS divided by the total power consumption at the BS. Since the EE of each link is non-cancave (which is a non-concave function over an affine function), characterizing this boundary is difficult. To meet this challenge, we relate this multi-cell MISO system to cognitive radio (CR) MISO channels by applying the concept of interference temperature (IT), and accordingly transform the EE boundary characterization problem into a set of fractional concave programming problems. Then, we apply the fractional concave programming technique to solve these fractional concave problems, and correspondingly give a parametrization for the EE boundary in terms of IT levels. Based on this characterization, we further present a decentralized algorithm to implement the multi-cell coordinated beamforming, which is shown by simulations to achieve the EE Pareto boundary.
\end{abstract}

\begin{IEEEkeywords}
multi-cell MISO systems, energy efficiency, Pareto boundary, coordinated beamforming.
\end{IEEEkeywords}

%
\IEEEpeerreviewmaketitle

\section{Introduction}
Green wireless communication has attracted a lot of interest recently due to the explosive growth of energy consumption of wireless communication networks and correspondingly increasing carbon dioxide (CO$_2$) emissions. In order to achieve this green goal, many innovative techniques among different layers of communication protocols have been proposed \cite{ref:Survey}. Among others, increasing the physical layer bits-per-Joule energy efficiency (EE), which is in general defined as the achievable data rates divided by the power consumption, has received increasing attention these days (see e.g. \cite{ref:Framework, ref:tradeoff, ref:XuJieTWC, ref:CooperativeIdling, ref:Sleeping} and references therein). In \cite{ref:Framework}, a general link level EE optimization framework is proposed based on fractional programming techniques. In \cite{ref:tradeoff}, the EE maximization of a downlink orthogonal frequency division multiple access (OFDMA) is addressed. In \cite{ref:XuJieTWC}, EE optimization for a multiple input multiple output (MIMO) broadcast channel is considered, where transmit covariance optimization and antenna selection technique are jointly applied to maximize the EE. In \cite{ref:CooperativeIdling, ref:Sleeping}, base station (BS) sleeping is employed to improve the EE of cooperative multi-cell wireless systems.

On the other hand, interference is the key factor limiting the performance of cellular networks in terms of both spectrum efficiency (SE) and EE. In order to mitigate the inter-cell interference, implementing BS cooperation is a promising solution. The BS cooperation can be mainly classified into two categories, namely joint processing and coordinated beamforming, where the former performs  symbol-level cooperation with both transmit message and channel state information (CSI) sharing, and the latter performs beamforming-level cooperation with CSI sharing only. In this paper, we focus on coordinated beamforming. There have been some works in the literature studying the optimal coordinated beamforming design (see e.g. \cite{ref:Complete, ref:Optimality, ref:Cooperative, ref:Optimal, ref:WeiYu} and references therein). In \cite{ref:WeiYu}, the transmit power minimization with minimum signal-to-interference-and-noise ratio (SINR) constraints is considered by applying uplink-downlink duality techniques. In \cite{ref:Complete, ref:Optimality, ref:Cooperative}, the Pareto boundary of the SE regions is characterized for multi-cell MISO systems. It is shown in \cite{ref:Complete} that the optimal beamforming vectors can be chosen as linear combinations of the zero-forcing (ZF) and the maximum-ratio transmission (MRT) beamformers, while in \cite{ref:Optimality, ref:Cooperative}, the SE Pareto boundary characterizing problem is solved by reformulating it as convex optimization problems. Furthermore, distributed algorithms to achieve the SE Pareto boundary are developed in \cite{ref:Cooperative, ref:Optimal}. However, these previous works all focus on the SE perspective. To our best knowledge, there are only \cite{ref:ICI-Aware, ref:MIMOInterf} considering energy efficient coordinated transmission in multi-cell systems. In \cite{ref:ICI-Aware}, a single-antenna OFDM multi-cell system is considered, for which an energy-efficient power optimization is developed based on non-cooperative games. In \cite{ref:MIMOInterf}, an energy efficient beamforming algorithm is derived for MIMO interference channels, for which local optimality is achieved.

In this paper, we investigate the optimal energy efficient coordinated beamforming for multi-cell MISO systems with $K$ multiple-antenna base stations (BS) and $K$ single-antenna mobile stations (MS), where each BS sends information to its own intended MS with cooperatively designed transmit beamforming. We assume single user detection at the MS by treating the interference as noise. By taking into account a realistic power consumption model at the BS, we characterize the Pareto boundary of the achievable EE region of the $K$ links, in which the EE of each link is defined as the achievable data rate at the MS divided by the total power consumption at the BS. Characterizing this boundary can reveal the fundamental EE tradeoffs among different links in interference networks, and thus is very important. Nonetheless, since the achievable date rate at each MS is a non-concave function due to the coupled mutual interference, the EE of each link is a non-concave function over an affine function, which makes characterizing this boundary very challenging. Fortunately, we find that through relating this multi-cell MISO system to cognitive radio (CR) channels by using the concept of interference temperature (IT), the original problem can be transformed into a set of parallel concave fractional programming problems. Then, by applying the fractional programming technique, we solve the reformulated problems globally optimally, and correspondingly develop a parameterized characterization of the EE Pareto boundary in terms of the IT levels. Based on this characterization, we further develop a distributed coordinated beamforming algorithm, which iteratively solves a set of
fractional programs. Through simulations, we show that this distributed algorithm can achieve the EE Pareto boundary for the multi-cell MISO systems.

Regarding the notations, for a square matrix $\bm{S}$, ${\rm Tr}(\bm{S})$, $\left| \bm{S} \right|$, $\bm{S}^{-1}$ and $\bm{S}^{1/2}$ denote the trace, determinant, inverse and square root of $\bm{S}$, respectively. $\bm{S}\succeq0$ means that $\bm{S}$ is positive semidefinite. For a matrix $\bm{X}$ of arbitrary size, $\bm{X}^H$ and $\bm{X}^T$  denote the conjugate transpose and transpose of $\bm{X}$, respectively. $\left\| \bm{x} \right\|$ denotes the Euclidean norm of a complex vector (scalar) $\bm{x}$. $\mathbb{C}^{m \times n}$ denotes the space of $m \times n$ complex matrices. $\mathbb{E}(\cdot)$ denotes the statistical expectation and the $\rm log(\cdot)$ function is with base 2 by default.

\section{System Model}
We consider a downlink multi-cell MISO system with $K$ BSs and $K$ MSs, where each BS serves its intended MS by cooperatively designed transmit beamforming. We assume that the $k$th BS is equipped with $M_k$ antennas, $M_k \ge 1$, and each MS is equipped with only one single antenna. Let $\bm{h}_{jk}^H \in \mathbb{C}^{1 \times M_k}$ denote the channel vector from BS $j$ to MS $k$, and $\bm{x}_k \in \mathbb{C}^{M_k \times 1}$ denote the transmitted signal of the $k$th BS, then the signal received by MS $k$ is given by
\begin{equation} \label{eq1}
\begin{array}{l}
y_k  = \bm{h}_{kk}^H \bm{x}_k  + \sum\limits_{j \ne k} {\bm{h}_{jk}^H \bm{x}_j  + z_k },
\end{array}
\end{equation}
where $z_k$ is the noise at the $k$th MS, which is modeled as a circularly symmetric complex Gaussian (CSCG) random variable with zero mean and variance $\sigma _k^2$, i.e., $z_k\sim \mathcal{CN}(0,\sigma _k^2)$.

We assume that the independent Gaussian codebook is used at each BS, i.e., $\bm{x}_k\sim \mathcal{CN}(0,\bm{S}_k)$,  where $\bm{S}_k = \mathbb{E} (\bm{x}_k\bm{x}_k^H)$ is the covariance of $\bm{x}_k$ with $\bm{S}_k\succeq0$, and $\bm{x}_k$'s are independent. For a given set of transmit covariance matrices of all BSs, $(\bm{S}_1, \ldots ,\bm{S}_K)$, and assuming that each MS performs single user detection with treating interference as noise, the achievable rate of the $k$th MS is given by
\begin{equation} \label{eq2}
\begin{array}{l}
\displaystyle R_k (\bm{S}_1 , \ldots ,\bm{S}_K ) = \log \left(1 + \frac{{\bm{h}_{kk}^H \bm{S}_k \bm{h}_{kk} }}{{\sum\nolimits_{j \ne k} {\bm{h}_{jk}^H \bm{S}_j \bm{h}_{jk}  + \sigma _k^2 } }} \right).
\end{array}
\end{equation}

On the other hand, we consider a realistic power consumption model for the BSs by taking into account a constant transmission-independent power representing the power consumed by air conditioner, data processing, circuits, etc. Suppose that the constant power is given by $P_c$, and the power amplifier efficiency is denoted as $\eta$.  Then the total power consumption at the $k$th BS is given as
\begin{equation} \label{eq3}
\begin{array}{l}
\displaystyle P_{{\rm total}, k} = \frac{{\rm{Tr}}(\bm{S}_k )}{\eta}  + P_c.
\end{array}
\end{equation}
Thus, we can define the EE of the $k$th link, i.e., the link between the $k$th BS and the $k$th MS, as the achievable rate at MS $k$ divided by the total power consumption of BS $k$:
\begin{equation} \label{eq4}
\begin{array}{l}
\displaystyle \xi _k (\bm{S}_1 , \ldots ,\bm{S}_K ) = \frac{{\log \left(1 + \frac{{\bm{h}_{kk}^H \bm{S}_k \bm{h}_{kk} }}{{\sum\nolimits_{j \ne k} {\bm{h}_{jk}^H \bm{S}_j \bm{h}_{jk}  + \sigma _k^2 } }}\right)}}{\frac{{\rm{Tr}}(\bm{S}_k )}{\eta}  + P_c}.
\end{array}
\end{equation}
Accordingly, we define the achievable EE region as the set of all EE tuples that can be achieved under a set of power constraints, denoted by $(P_1, \ldots, P_K)$:
\begin{equation} \label{eq5}
\begin{array}{l}
{\mathcal{E}} \buildrel \Delta \over = \bigcup\limits_{\{ \bm{S}_k \} :{\rm{Tr}}(\bm{S}_k ) \le P_k ,k = 1, \ldots ,K} {\{ (\varepsilon _1 , \ldots ,\varepsilon _K ):}  \\
{{\rm{ 0}} \le \varepsilon _k  \le \xi _k (\bm{S}_1 , \ldots ,\bm{S}_K ) ,k = 1,2, \ldots ,K\} }.
\end{array}
\end{equation}
In this paper, our objective is to characterize the EE Pareto boundary or find the Pareto optimal EE tuples, which is the outer boundary of the EE region given in (\ref{eq5}). More precisely, we can define the Pareto optimality of the EE as follows:

\emph{Definition 1}: An EE tuple $(\varepsilon _1 , \ldots ,\varepsilon _K )$ is Pareto optimal if there is no other tuples $(\varepsilon _1', \ldots ,\varepsilon _K')$ satisfying $(\varepsilon _1' , \ldots ,\varepsilon _K' ) \ge (\varepsilon _1 , \ldots ,\varepsilon _K )$ and $(\varepsilon _1' , \ldots ,\varepsilon _K' ) \ne (\varepsilon _1 , \ldots ,\varepsilon _K )$ (the inequalities are component-wise) at the same time.

The EE Pareto boundary can describe the achievable EE performance upper bound, reveal the fundamental EE tradeoffs among different links in a interference network, and thus is very important. However, it can be observed that the EE in (\ref{eq4}) is a non-concave function divided by an affine function, and thus is non-concave. Therefore, characterizing this EE Pareto boundary is non-trivial.

\section{A Parametrization of the EE Pareto Boundary for the Multi-cell MISO System}
In this section, we characterize the EE Pareto boundary by reformulating it as a set of concave fractional programs. First, we introduce auxiliary variables $\Gamma _{kj}$, $k = 1, \ldots, K, j = 1, \ldots, K, j \ne k$, where $\Gamma _{kj}$ denotes the maximum tolerable interference from the $k$th BS to the $j$th MS with $\Gamma _{kj} \ge 0$, namely the IT constraint. For notational convenience, let $\bm{\Gamma}$ be the vector consisting of all $K(K-1)$  different $\Gamma _{kj}$'s, and $\bm{\Gamma}_k$ be the vector consisting of all $2(K-1)$ different $\Gamma _{kj}$'s and $\Gamma _{jk}$'s, $j = 1, \ldots ,K,j \ne k$, for any given $k \in \{1, \ldots, K\}$.

With these auxiliary variables, we define $K$ parallel transmit covariance optimization problems, each for one of the $K$ BSs expressed as
\begin{eqnarray}
& \displaystyle \mathop {{\rm{max}}}\limits_{\bm{S}_k}  \frac{{\log \left(1 + \frac{{\bm{h}_{kk}^H \bm{S}_k \bm{h}_{kk} }}{{\sum\nolimits_{j \ne k} {\Gamma _{jk}  + \sigma _k^2 } }}\right)}}{{\frac{{\rm{Tr}}(\bm{S}_k )}{\eta}  + P_c }}  \label{eq6} \\
& \mathrm{s.t.} \ {\rm{     }}\bm{h}_{kj}^H \bm{S}_k \bm{h}_{kj}  \le \Gamma _{kj} ,\forall j \ne k  \label{constraint1} \\
& \ {\rm{Tr}}(\bm{S}_k ) \le P_k ,\bm{S}_k  \succeq 0, \label{constraint2}
\end{eqnarray}
where $k \in \{1, \ldots, K\}$. Note that in the above problem $\bm{\Gamma}_k$ is fixed for a given $k$, and $\bm{S}_k$ is a general-rank transmit covariance matrix. For notational convenience, we denote $\mathbb{S}_k$ as the feasible region of the transmit covariance $\bm{S}_k$,  which is specified by the constraints in (\ref{constraint1}) and (\ref{constraint2}). We also denote $E_k (\bm{\Gamma}_k)$ as the optimal objective value of problem (\ref{eq6}).

It can be easily verified that the feasible region $\mathbb{S}_k$ is convex, and the objective of (\ref{eq6}) is a concave function over an affine function, thus problem (\ref{eq6}) is a concave fractional program \cite{ref:Fractional}. Moreover, from [16], it also follows that the objective function in (\ref{eq6}) is pseudo-concave, so the maximum for problem (\ref{eq6}) is unique, i.e., any local optimal point is globally optimal. In order to solve problem (\ref{eq6}), we apply the technique of concave fractional programming \cite{ref:Fractional} and relate it to a parametric convex program by separating the numerator and denominator of the objective function. By introducing a parameter $\gamma_k$, we can define the parametric convex program as :
\begin{eqnarray} \label{axeq1}
F{\rm{(}}\gamma _k {\rm{)}} =
& \mathop {{\rm{max}}}\limits_{\bm{S}_k \in \mathbb{S}_k} & \displaystyle \log \left(1 + \frac{{\bm{h}_{kk}^H \bm{S}_k \bm{h}_{kk} }}{{\sum\nolimits_{j \ne k} {\Gamma _{jk}  + \sigma _k^2 } }}\right) \nonumber\\
& & \displaystyle - \gamma _k \left(\frac{{\rm{Tr}}(\bm{S}_k )}{\eta}  + P_c \right){\rm{     }},
\end{eqnarray}
which can be easily verified to be a strictly decreasing, continuous function on $\gamma _k$. Suppose that the unique solution of $F{\rm{(}}\gamma _k {\rm{)}} = 0$ is denoted by $\gamma_k^*$, then according to \cite{ref:Fractional}, problem (\ref{eq6}) can be equivalently solved via solving the problem in (\ref{axeq1}) with $\gamma_k = \gamma_k^*$, and moreover, the optimal objective value $E_k (\bm{\Gamma}_k)$ of problem (\ref{eq6}) is equal to $\gamma_k^*$. Therefore, in the following, we should first solve the problem in (\ref{axeq1}) for any given $\gamma_k$, and then search over $\gamma_k$ to find $\gamma_k^*$ with $F(\gamma_k^*) = 0$.

For a fixed $\gamma_k$, the problem in (\ref{axeq1}) is a convex optimization problem, so we apply the Lagrange duality method \cite{ref:Convex} to solve it. Let $\lambda_{kj}, j \ne k$, and $\lambda_{kk}$ be the non-negative dual variables for the problem in (\ref{axeq1}), associated with the $k$th BS's IT constraint for the $j$th MS and its own transmit power constraint, respectively. Denote $\bm{\lambda} _k$ as a vector consisting of all ${\lambda}_{kj}$'s and ${\lambda}_{kk}$. The Lagrangian of the problem in (\ref{axeq1}) is expressed as
\begin{eqnarray} \label{axeq3}
\displaystyle L(\bm{S}_k ,\bm{\lambda} _k ) & = & \log \left(1 + \frac{{\bm{h}_{kk}^H \bm{S}_k \bm{h}_{kk} }}{{\sum\nolimits_{j \ne k} {\Gamma _{jk}  + \sigma _k^2 } }}\right)    \nonumber\\
& & - \gamma _k \left(\frac{{\rm{Tr}}(\bm{S}_k )}{\eta} + P_c \right) - \lambda _{kk} {\rm{(Tr}}(\bm{S}_k ) - P_k {\rm{)}} \nonumber\\
& & \displaystyle - \sum\limits_{j \ne k} {\lambda _{kj} {\rm{ (}}\bm{h}_{kj}^H \bm{S}_k \bm{h}_{kj}  - \Gamma _{kj} )} .
\end{eqnarray}
Then the dual function is given by
\begin{equation} \label{axeq4}
\begin{array}{l}
g(\bm{\lambda} _k ) = \mathop {\max }\limits_{{\bf{S}}_k  \succeq 0} L({\bm{S}}_k ,\bm{\lambda} _k ).
\end{array}
\end{equation}
Accordingly, the dual problem is defined as
\begin{equation} \label{axeq5}
\begin{array}{l}
\mathop {\rm{min} }\limits_{{\bm{\lambda }}_k \ge 0} g(\bm{\lambda} _k ),
\end{array}
\end{equation}
where $\bm{\lambda}_k \ge 0$ means component-wise non-negative. Since the problem in (\ref{axeq1}) is convex and satisfies the slater's condition \cite{ref:Convex}, the duality gap between the problem in (\ref{axeq1}) and its dual problem (\ref{axeq5}) is zero; thus, the problem in (\ref{axeq1}) can be equivalently solved via solving (\ref{axeq5}).

First, we obtain the dual function $g(\bm{\lambda}_k)$ in (11) under given $\bm{\lambda}_k$ by solving the maximization problem as follows (by discarding irrelevant constant terms):
\begin{equation} \label{axeq6}
\begin{array}{l}
\displaystyle \mathop {{\rm{max}}}\limits_{\bm{S}_k \succeq 0} \log \left(1 + \frac{{\bm{h}_{kk}^H \bm{S}_k \bm{h}_{kk} }}{{\sum\nolimits_{j \ne k} {\Gamma _{jk}  + \sigma _k^2 } }}\right) - {\rm{Tr}}(\bm{B}_k \bm{S}_k ),
\end{array}
\end{equation}
where $\bm{B}_k  \buildrel \Delta \over = \sum\limits_{j \ne k} {\lambda _{kj} \bm{h}_{kj} \bm{h}_{kj}^H  + (\lambda _{kk}  + \frac{{\gamma _k }}{\eta })} \bm{I}$ and $\bm{B}_k \succeq 0$ of dimension $M_k \times M_k$.
Motivated by \cite[Appendix I]{ref:Cooperative}, the closed-form solution of problem (\ref{axeq6}) is given by
\begin{equation} \label{axeq7}
\begin{array}{l}
\bm{S}_k^{\star}   = \frac{{\left( {\frac{1}{{{\rm{ln}}2}} - \frac{{\sum\nolimits_{j \ne k} {\Gamma _{jk}  + \sigma _k^2 } }}{{\left\| {\bm{h}_{kk}^H \bm{B}_k ^{ - 1/2} } \right\|^2 }}} \right)^ +  }}{{\left\| {\bm{B}_k ^{ - 1/2} \bm{h}_{kk} } \right\|^2 }}\bm{B}_k ^{ - 1} \bm{h}_{kk} \bm{h}_{kk}^H \bm{B}_k ^{ - 1},
\end{array}
\end{equation}
where $(x)^+\buildrel \Delta \over=\max (0,x)$. With the obtained dual function $g(\bm{\lambda} _k )$ for any given $\bm{\lambda} _k$, we can then solve the dual problem (\ref{axeq5}). Since $g(\bm{\lambda}_k)$ is convex but not necessarily differentiable, the dual problem (\ref{axeq5}) can be solved by the subgradient based method such as ellipsoid method \cite{ref:ellipsoid}, by using the fact that the subgradients of $g(\bm{\lambda}_k)$ are $\Gamma _{kj}  - \bm{h}_{kj}^H \bm{S}_k^{\star} \bm{h}_{kj}$ and $P_k  - {\rm{Tr}}(\bm{S}_k^{\star})$ for $\lambda_{kj}$, $k \neq j$, and $\lambda_{kk}$, respectively. Suppose that the optimal solution for the dual problem (\ref{axeq5}) is given by $\lambda_{kj}^*, j \ne k$, and $\lambda_{kk}^*$, then the corresponding $\bm{S}_k^{\star}$ becomes the optimal solution of the problem in (\ref{axeq1}), denoted by $\bm{S}_k^*$. Note that $\bm{S}_k^{\star}$ in (\ref{axeq7}) is in general a rank-one matrix, so we can give the following proposition to denote the optimal solution of the problem in (\ref{axeq1}) based on the above derivation.

\emph{Proposition 1:} The optimal solution of the problem in (\ref{axeq1}) is rank-one, i.e. $\bm{S}_k^*=\bm{w}_k\bm{w}_k^H$, and
\begin{equation} \label{eq9}
\begin{array}{l}
\bm{w}_k = \displaystyle \sqrt {p_k } \bm{B}_k ^{* - 1} \bm{h}_{kk},
\end{array}
\end{equation}
with $p_k$ being
\begin{equation} \label{eq10}
\begin{array}{l}
\displaystyle p_k  =  \left( {\frac{1}{{{\rm{ln}}2}} - \frac{{\sum\nolimits_{j \ne k} {\Gamma _{jk}  + \sigma _k^2 } }}{{\left\| {\bm{h}_{kk}^H \bm{B}_k ^{* - 1/2} } \right\|^2 }}} \right)^ +  \frac{1}{{\left\| {\bm{B}_k ^{* - 1/2} \bm{h}_{kk} } \right\|^2 }},
\end{array}
\end{equation}
where $\bm{B}_k^{*}  \buildrel \Delta \over = \sum\limits_{j \ne k} {\lambda _{kj}^* \bm{h}_{kj} \bm{h}_{kj}^H  + (\lambda _{kk}^*  + \frac{{\gamma _k }}{\eta })} \bm{I}$. $\lambda_{kj}^*, j \ne k$, and $\lambda_{kk}^*$ are the optimal solution of the dual problem (\ref{axeq5}).

As the solution $\bm{S}_k^*$ and the optimal objective value $F{\rm{(}}\gamma _k {\rm{)}}$ of the problem in (\ref{axeq1}) can be obtained for any given $\gamma_k$, a simple bisection method can be applied to find $\gamma_k^*$. With obtained $\gamma_k^*$, the corresponding optimal $\bm{S}_k^*$ becomes the solution of problem (\ref{eq6}). Thus, problem (\ref{eq6}) is efficiently solved. We summarize the iterative algorithm to solve problem (\ref{eq6}) in Algorithm \ref{IteWater}.
\begin{algorithm}
\algsetup{linenosize=\small}
\small
\caption{Iterative Algorithm for Solving Problem (\ref{eq6})}\label{IteWater}
\begin{algorithmic}
\STATE Initialize $\gamma_{k}^a>0$ and $\gamma_{k}^b>0$ so that $F{\rm{(}}\gamma _k^a {\rm{)}} > 0$ and $F{\rm{(}}\gamma _k^b {\rm{)}} < 0$.
\REPEAT
\STATE Set $\gamma_{k}^* = \frac{1}{2}(\gamma_{k}^a+\gamma_{k}^b)$.
\STATE Solve the problem in (\ref{axeq1}) with $\gamma_{k} = \gamma_{k}^*$ by applying Proposition 1, and accordingly obtain the optimal objective value $F{\rm{(}}\gamma _k^* {\rm{)}}$.
\IF{$F(\gamma _k^* ) > 0$}
\STATE $\gamma _k^a \gets \gamma _k^*$
\ELSE
\STATE $\gamma _k^b \gets \gamma _k^*$
\ENDIF
\UNTIL{$| \gamma_{k}^a-\gamma_{k}^b | \le \epsilon$}
\end{algorithmic}
\end{algorithm}

Next, the following proposition identifies the relationship between problem (\ref{eq6}) and the EE Pareto boundary, and accordingly a parameterized characterization of the EE Pareto boundary is given in terms of $\bm{\Gamma}$.

\emph{Proposition 2:} For any EE tuple $(\xi _1 , \ldots ,\xi _K )$ on the EE Pareto boundary of the EE region defined in (\ref{eq5}), which is achievable with a set of transmit covariance matrices, $(\bm{S}_1 , \ldots ,\bm{S}_K )$, there is a corresponding IT constraint vector, $\bm{\Gamma}  \ge 0$, with $\Gamma _{kj}  = {\rm{ }}\bm{h}_{kj}^H \bm{S}_k \bm{h}_{kj}, {\rm{ }}\forall j \ne k, j = 1, \ldots , K,$ and $k \in \{1, \ldots, K\}$, such that $\xi_k = E_k(\bm{\Gamma}_k), \forall k$, and $\bm{S}_k$ is the optimal solution of Problem (\ref{eq6}) for the given $k$.
\begin{proof}
See Appendix I.
\end{proof}

Combining Propositions 1 and 2, it follows that beamforming is indeed optimal to achieve any Pareto optimal EE tuples, which is similar to the SE case \cite{ref:Cooperative}. Moreover, it is also shown that the Pareto boundary can be characterized by $K(K-1)$ real parameters, i.e., the parameters given in $\bm{\Gamma}$.

It is interesting to consider a special case with  $P_c=0$, which corressponds to the case that the transmit power dominates the total power consumption. In this case, it is easy to verify that the EE boundary is achieved by $\bm{S}_k^*=\bm{w}_k\bm{w}_k^H$, where
\begin{equation} \label{OptNoPcir}
\begin{array}{l}
\displaystyle \bm{w}_k =  \sqrt{p_k} \frac {\bm{h}_{kk}} {\left\| \bm{h}_{kk} \right\|}
\end{array}
\end{equation}
and $p_k\to0$. From (\ref{OptNoPcir}), we can see that the Pareto boundary point with $P_c=0$ is achieved by MRT beamforming as well as extremely small transmit power. Intuitively, transmitting at extremely low power minimizes the interference and accordingly maximizes the EE. In this case, the optimal EE tuple of the boundary point can be given as
\begin{equation} \label{eq11}
\begin{array}{l}
\displaystyle (\eta\frac{\left\|\bm{h}_{11}\right\|^2}{\sigma _1^2\rm{ln}2},\eta\frac{\left\|\bm{h}_{22}\right\|^2}{\sigma _2^2\rm{ln}2}, \ldots, \eta\frac{\left\|\bm{h}_{KK}\right\|^2}{\sigma _K^2\rm{ln}2}).
\end{array}
\end{equation}
Note that this phenomenon is consistent with the single cell SISO case with $P_c = 0$, for which it is shown in \cite{ref:EEwoPcir} that transmitting with extremely low power achieves the optimal EE.

In order to demonstrate the effect of the constant power, Fig.1 shows the EE region with different values of the constant power, for which the parameters are given in Section V. It is observed that as $P_c$ increases, the optimal EE decreases, and accordingly, the EE region with large $P_c$ is smaller than that with small $P_c$. In the case of $P_c=0$, we can see that the Pareto boundary consists of only one Pareto optimal point which is given in (\ref{eq11}), and thus the whole region is a box. In the case of $P_c > 0$, we can see that the EE Pareto boundary consists of multiple boundary points, which reveals the EE tradeoffs among different links in the multi-cell MISO system.
\begin{figure}[!t]
\begin{center}
\includegraphics[height = 2.5in] {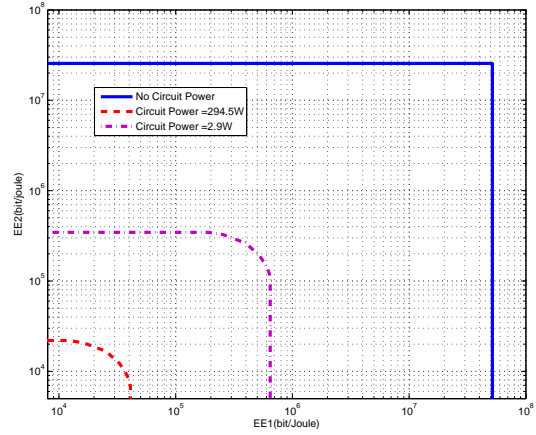}
\end{center}
\caption{EE Regions with different constant power $P_c$.} \label{fig2}
\end{figure}

\section{Decentralized Algorithm for Multi-Cell Cooperative Beamforming}
In this section, we present a decentralized iterative algorithm to implement the multi-cell coordinated beamforming for achieving the EE Pareto boundary, for which in each iteration step, concave fractional program (\ref{eq6}) is solved with given IT constraints ${\Gamma}_{kj}$'s and ${\Gamma}_{jk}$'s.

The procedure of our proposed algorithm is described as follows:
First, the BSs set their beamforming vectors individually through solving problems in (\ref{eq6}) with a set of prescribed IT constraints, for which Algorithm \ref{IteWater} is applied. Then BS pair $(i,j)$, $\forall i,j \in \{1, \ldots, K\}$, $i \ne j$, communicates with each other for updating their mutual IT constraints based on the following rules:
\begin{equation} \label{eq12}
\begin{array}{l}
[\Gamma '_{ij} ,\Gamma '_{ji} ]^T  = [\Gamma _{ij} ,\Gamma _{ji} ]^T  + \delta _{ij}  \cdot \bm{d}_{ij},
\end{array}
\end{equation}
where $\delta _{ij}$ is a small step-size, $\bm{d}_{ij}$ is any vector that satisfies $\bm{D}_{ij} \bm{d}_{ij}  > 0$ with $\bm{D}_{ij}$ given as
\begin{equation} \label{eq13}
\begin{array}{l}
\bm{D}_{ij}  = \begin{bmatrix}
   {\frac{{\partial E_i (\bm{\Gamma} _i )}}{{\partial \Gamma _{ij} }}} & {\frac{{\partial E_i (\bm{\Gamma} _i )}}{{\partial \Gamma _{ji} }}}  \\
   {\frac{{\partial E_j (\bm{\Gamma} _j )}}{{\partial \Gamma _{ij} }}} & {\frac{{\partial E_j (\bm{\Gamma} _j )}}{{\partial \Gamma _{ji} }}}  \\
\end{bmatrix},
\end{array}
\end{equation}
in which $E_i (\bm{\Gamma} _i )$ and $E_j (\bm{\Gamma} _j )$ denote the optimal value of problem (\ref{eq6}) for BS $i$ and BS $j$. Furthermore, if we denote $\bm{D}_{ij} = \begin{bmatrix} a & b \\ c & d \\ \end{bmatrix}$, then an example for $\bm{d}_{ij}$ can be shown as \cite{ref:Cooperative}
\begin{equation} \label{eq14}
\begin{array}{l}
\bm{d}_{ij}  = sign(ad - bc) \cdot [\alpha _{ij} d - b,a - \alpha _{ij} c]^T ,{\rm{ }}\alpha _{ij}  \ge 0.
\end{array}
\end{equation}
With updated $\Gamma '_{ij}$ and $\Gamma '_{ji}$, the above procedures are iteratively implemented until $\left| {\bm{D}_{ij} } \right| = 0$. Note that $\left| {\bm{D}_{ij} } \right| = 0$ is a necessary condition for guaranteeing that the vector $\bm{\Gamma}$ corresponds to the EE Pareto boundary, as given in the following proposition.

\emph{Proposition 3:} For an arbitrarily chosen $\bm{\Gamma}>0$, if the optimal values of the problems in (\ref{eq6}) for all $k$'s, $E_k (\bm{\Gamma}_k)$'s, are Pareto optimal on the boundary of the EE region for the multi-cell MISO system, then for any BS pair $(i,j)$,  it must holds that $\left| {\bm{D}_{ij} } \right| = 0$.

The proof of Proposition 3 follows \cite[Appendix III]{ref:Cooperative} directly and thus is omitted here. Note that similar to the SE case as in \cite{ref:Cooperative}, although the sufficiency of the Pareto optimality is not proved, simulation results will be given in the next section to verify the performance of our proposed algorithm.

To complete the above algorithm, we show how to calculate $\bm{D}_{ij}$  in each iteration in the following. For the $i$th BS, suppose that the optimal solution of the problem (\ref{eq6}) is given as $\bm{S}^*_i$. Then, based on the sensitivity results of concave fractional programming\cite[Section 4]{ref:Fractional}, we have
\begin{equation} \label{eq15}
\begin{array}{l}
\displaystyle \frac{{\partial E_i (\bm{\Gamma} _i )}}{{\partial \Gamma _{ij} }} = \frac{{\lambda _{ij}^* }}{{\frac{{\rm{Tr}}(\bm{S}_i^*  )}{\eta}  + P_c }},
\end{array}
\end{equation}
where $\lambda _{ij}^*$ is the optimal Lagrange dual variable corresponding to the IT constraint between the $i$th BS and the $j$th MS of problem (\ref{axeq1}) with $\gamma_i=\gamma _i^*$.

On the other hand, by calculating the derivative of the objective function in (\ref{eq6}), we can obtain
\begin{equation} \label{eq16}
\begin{array}{l}
\displaystyle \frac{{\partial E_i (\bm{\Gamma} _i )}}{{\partial \Gamma _{ji} }} = \displaystyle \frac{- \bm{h}_{ii}^H  \bm{S}_i^*  \bm{h}_{ii}\left(\frac{{\rm{Tr}}(\bm{S}_i^* )}{\eta}  + P_c\right)^{-1}}{{(\sum\nolimits_{l \ne i} {\Gamma _{li}  + \sigma _i^2 } )(\sum\nolimits_{l \ne i} {\Gamma _{li}  + \sigma _i^2  + \bm{h}_{ii}^H  \bm{S}_i^*  \bm{h}_{ii} } )}\ln 2}.
\end{array}
\end{equation}
Similarly, $(\partial E_j(\bm{\Gamma} _j ))/(\partial {\Gamma} _{ij})$ and $(\partial E_j(\bm{\Gamma} _j ))/(\partial \Gamma _{ji})$ can be obtained by solving problem (\ref{eq6}) with $k=j$. Thus, $\bm{D}_{ij}$ is obtained.

To summarize, the distributed algorithm can be given in Algorithm \ref{Beamforming}.
\begin{algorithm}
\algsetup{linenosize=\small}
\small
\caption{\ \protect\\ Distributed Energy Efficient Coordinated Beamforming}\label{Beamforming}
\begin{algorithmic}
\STATE Initialize $\bm{\Gamma} \geq \bm{0}$ over the whole network.
\STATE BS $k$ sets $\bm{w}_k$ via solving problem (\ref{eq6}) with the initialized $\bm{\Gamma}_k$, $k=1, \ldots, K$.
\REPEAT
\FOR {$i=1, \ldots, K$, $j=1, \ldots, K$, $i \ne j$,}
\STATE BS $i$ computes $\frac{{\partial E_i (\bm{\Gamma} _i )}}{{\partial \Gamma _{ij} }}$ and $\frac{{\partial E_i (\bm{\Gamma} _i )}}{{\partial \Gamma _{ji} }}$ with the given $\bm{\Gamma}_i$.
\STATE BS $j$ computes $\frac{{\partial E_j (\bm{\Gamma} _j )}}{{\partial \Gamma _{ij} }}$ and $\frac{{\partial E_j (\bm{\Gamma} _j )}}{{\partial \Gamma _{ji} }}$ with the given $\bm{\Gamma}_j$.
\STATE BS $i$ and BS $j$ send their computation results to each other.
\STATE BS $i$ ($j$) computes $\bm{d}_{ij}$, then update $\Gamma_{ij}$ and $\Gamma_{ji}$ (cf. (\ref{eq13})).
\STATE BS $i$ ($j$) resets $\bm{w}_k$ via solving problem (\ref{eq6}) with the updated $\bm{\Gamma}_i$ ($\bm{\Gamma}_j$).
\ENDFOR
\UNTIL{$\left| {\bm{D}_{ij} } \right| = 0$, $\forall i \ne j$.}
\end{algorithmic}
\end{algorithm}

\section{Simulation Result}
In this section, we present numerical results to validate our theoretic results. We consider a two-cell MISO system with a bandwidth of $5\rm{MHz}$, where $K=2$, $M_1=M_2=3$, , $\eta = 0.38$ and $\sigma _1^2=\sigma _2^2=-110 \rm{dBm}$. In order to obtain the Pareto boundary, we solve the problems in (\ref{eq6}) for $k=1,2$, with different pairs of $\Gamma_{12}$ and $\Gamma_{21}$ within their respective ranges, and then take a closure operation over all the obtained EE pairs.

\begin{figure}[h]
\begin{center}
\includegraphics[height = 2.5in] {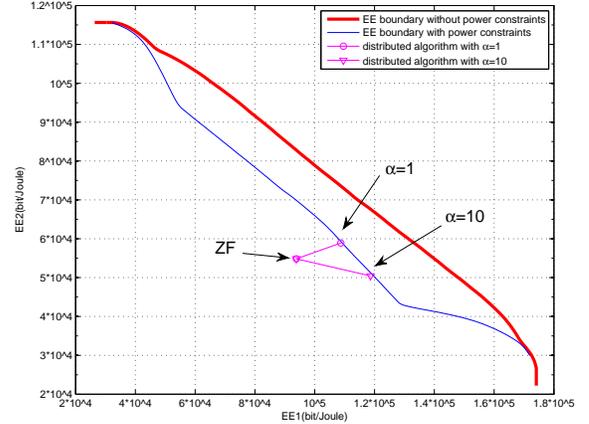}
\end{center}
\caption{EE Pareto boundary of the multi-cell MISO system with realistic power model.} \label{fig1}
\end{figure}

In Fig. 2, we set $P_c = 294.5W$, and demonstrate the EE Pareto boundary for this system. We consider two cases, i.e., with and without power constraints. For the case with power constraints, we set that $ P_1=P_2=43\rm{dBm}$, while for the case without power constraints, it is equivalent to $P_1, P_2 \to \infty$. It is observed that the EE region without power constraints is larger than that with power constraints, which is due to the fact that the former case has a larger feasible transmit power region. It is also observed that the EE boundary is bounded for the case without transmit power constraints. This is because that the optimal EE is always attained with limited transmit power, which is different from the case of SE maximization. In Fig.2, we also demonstrate the effectiveness of the distributed energy efficient coordinated beamforming algorithm. In the simulation, we use the EE-optimal ZF beamforming scheme as the initial point, and implement Algorithm 2 with different values of $\alpha$, where $\alpha = \alpha_{12} = \alpha_{21}$. It is observed that the EE tuple converges to EE Pareto-optimal pairs, which shows the efficiency of our distributed algorithm.

\section{Conclusion}
In this paper, we characterize EE Pareto boundary for a multi-cell MISO system with a realistic power model at the BS, in order to reveal the fundamental EE tradeoffs among different links in interference networks. Through utilizing the concept of IT and applying concave fractional programming techniques, we develop a parameterized characterization of the EE Pareto boundary in terms of IT levels. Based on this characterization, we further develop a distributed coordinated beamforming algorithm, which is verified by simulations to achieve the EE Pareto boundary.


%

\appendices

\section*{Appendix I}
First, assuming that a given set of $(\bm{S}_1 , \ldots ,\bm{S}_K )$ achieves the Pareto-optimal EE tuple $(\xi _1 , \ldots ,\xi _K )$ for the multi-cell MISO system, it then follows that for any $k  \in \{ 1, \ldots, K \}$
\begin{equation} \label{axeq8}
\begin{array}{l}
\displaystyle \xi _k  = \frac{{\log \left(1 + \frac{{\bm{h}_{kk}^H \bm{S}_k \bm{h}_{kk} }}{{\sum\nolimits_{j \ne k} {\bm{h}_{jk}^H \bm{S}_j \bm{h}_{jk}  + \sigma _k^2 } }}\right)}}{{\frac{{\rm{Tr}}(\bm{S}_k )}{\eta}  + P_c }}.
\end{array}
\end{equation}
Since $\Gamma _{jk}  = {\rm{ }}\bm{h}_{jk}^H \bm{S}_j \bm{h}_{jk} ,{\rm{ }}\forall j \ne k$, (\ref{axeq8}) can be rewritten as
\begin{equation} \label{axeq9}
\begin{array}{l}
\displaystyle \xi _k  = \frac{{\log \left(1 + \frac{{\bm{h}_{kk}^H \bm{S}_k \bm{h}_{kk} }}{{\sum\nolimits_{j \ne k} {\Gamma _{jk}  + \sigma _k^2 } }}\right)}}{{\frac{{\rm{Tr}}(\bm{S}_k )}{\eta}  + P_c }}.
\end{array}
\end{equation}
Note that (\ref{axeq9}) has the same form with the objective function of problem (\ref{eq6}). Furthermore, we have that ${\rm{Tr}}(\bm{S}_k ) \le P_k$ and $\Gamma _{kj}  = {\rm{ }}\bm{h}_{kj}^H \bm{S}_k \bm{h}_{kj} ,{\rm{ }}\forall j \ne k$, so $\bm{S}_k$ satisfies the constraints given in problem (\ref{eq6}) for any given $k$. Therefore, the Pareto-optimal covariance matrix $\bm{S}_k$ is a feasible solution for problem (\ref{eq6}).

Next, we use contradiction to prove that $\bm{S}_k$ is indeed the optimal solution of Problem (\ref{eq6}) and $\xi_k = E_k (\bm{\Gamma} _k )$ for any given $k$. Suppose the solution for problem (\ref{eq6}), denoted by $\bm{S}_k^\star$, is not equal to $\bm{S}_k$ for a given $k$. According to \cite{ref:Fractional}, we have
\begin{eqnarray} \label{axeq11}
\xi _k & = & \displaystyle \frac{{\log \left(1 + \frac{{\bm{h}_{kk}^H \bm{S}_k \bm{h}_{kk} }}{{\sum\nolimits_{j \ne k} {\Gamma _{jk}  + \sigma _k^2 } }}\right)}}{{\frac{{\rm{Tr}}(\bm{S}_k )}{\eta}  + P_c }} \nonumber\\
& < & \gamma^* = \displaystyle \frac{{\log \left(1 + \frac{{\bm{h}_{kk}^H \bm{S}_k^\star \bm{h}_{kk} }}{{\sum\nolimits_{j \ne k} {\Gamma _{jk}  + \sigma _k^2 } }}\right)}}{{\frac{{\rm{Tr}}(\bm{S}_k^\star )}{\eta}  + P_c }} \nonumber\\
& = & \displaystyle \frac{{\log \left(1 + \frac{{\bm{h}_{kk}^H \bm{S}_k^\star h_{kk} }}{{\sum\nolimits_{j \ne k} {\bm{h}_{jk}^H \bm{S}_j \bm{h}_{jk}  + \sigma _k^2 } }}\right)}}{{\frac{{\rm{Tr}}(\bm{S}_k^\star )}{\eta}  + P_c }} \buildrel \Delta \over = \varepsilon _k.
\end{eqnarray}
Furthermore, since $\bm{h}_{kj}^H \bm{S}_k^\star \bm{h}_{kj}  \le \Gamma _{kj} ,{\rm{ }}\forall j \ne k$, we have for any $j \ne k$,
\begin{eqnarray} \label{axeq12}
\xi _j  & = &  \displaystyle \frac{{\log \left(1 + \frac{{\bm{h}_{jj}^H \bm{S}_j \bm{h}_{jj} }}{{\sum\nolimits_{i \ne j} {\Gamma _{ij}  + \sigma _j^2 } }}\right)}}{{\frac{{\rm{Tr}}(\bm{S}_j )}{\eta}  + P_c }} \nonumber\\
& \le & \displaystyle \frac{{\log \left(1 + \frac{{\bm{h}_{jj}^H \bm{S}_j \bm{h}_{jj} }}{{\sum\nolimits_{i \ne j,k} {\Gamma _{ij}  + \bm{h}_{kj}^H \bm{S}_k^\star \bm{h}_{kj}  + \sigma _j^2 } }}\right)}}{{\frac{{\rm{Tr}}(\bm{S}_j )}{\eta}  + P_c }} \nonumber\\
& = & \displaystyle \frac{{\log \left(1 + \frac{{\bm{h}_{jj}^H \bm{S}_j \bm{h}_{jj} }}{{\sum\nolimits_{i \ne j,k} {\bm{h}_{ij}^H \bm{S}_j \bm{h}_{ij}  + \bm{h}_{kj}^H \bm{S}_k^\star \bm{h}_{kj}  + \sigma _j^2 } }}\right)}}{{\frac{{\rm{Tr}}(\bm{S}_j )}{\eta}  + P_c }} \nonumber\\
& \buildrel \Delta \over = & \varepsilon _j.
\end{eqnarray}
Thus, for another set of transmit covariance matrices given by $(\bm{S}_1 , \ldots ,\bm{S}_{k - 1} ,\bm{S}_k^\star  ,\bm{S}_{k + 1} , \ldots ,\bm{S}_K )$, the corresponding achievable EE tuple for the multi-cell MISO system, $(\varepsilon _1 , \ldots ,\varepsilon _K )$, satisfies that $\xi _k  < \varepsilon _k$ and $\xi _j  \le \varepsilon _j ,{\rm{ }}\forall j \ne k$, which contradicts the fact that $(\xi _1 , \ldots ,\xi _K )$ is a Pareto-optimal EE tuple for the system. Hence, the presumption cannot be true. Accordingly, we have  $\bm{S}_k^\star  = \bm{S}_k $ and $\xi_k = E_k (\bm{\Gamma} _k )$.

\ifCLASSOPTIONcaptionsoff
  \newpage
\fi

%









\begin{thebibliography}{8}
\bibitem{ref:Survey}
C. Yan, S. Zhang, S. Xu, and G. Y. Li, ``Fundamental trade-offs on green wireless networks'', \emph{IEEE Commun. Mag.}, vol. 49, no. 6, pp. 30-37, Jun. 2011.

\bibitem{ref:Framework}
C. Isheden, Z. Chong, E. Jorswieck, and G. Fettweis, ``Framework for link-level energy efficiency optimization with informed transmitter,'' \emph{IEEE Trans.  Wireless Commun.}, vol. 11, no. 8, pp. 2946-2957, Aug. 2012.

\bibitem{ref:tradeoff}
C. Xiong, G. Y. Li, S. Zhang, Y. Chen, and S. Xu, ``Energy- and spectral efficiency tradeoff in downlink OFDMA networks,'' \emph{IEEE Trans. Wireless Commun.}, vol. 10, no. 1, pp. 3874-3886, Nov. 2011.

\bibitem{ref:XuJieTWC}
J. Xu and L. Qiu, ``Energy efficiency optimization for MIMO broadcast channels,'' \emph{IEEE Trans. Wireless Commun.}, vol. 12, no. 2, pp. 690-701, Feb. 2013.

\bibitem{ref:CooperativeIdling}
J. Xu, L. Qiu, and C. Yu, ``Improving network energy efficiency through cooperative idling in the multi-cell systems,'' \emph{EURASIP J. Wireless Commun. and Net.}, vol. 2011, no. 1, p. 165, 2011.

\bibitem{ref:Sleeping}
S. Q. Han, C. Y. Yang, G. Wang, and M. Lei, ``On the energy efficiency of base station sleeping with multicell cooperative transmission,'' in \emph{Proc. 2011 PIMRC}, pp.1536-1540, Sep. 2011.

\bibitem{ref:WeiYu}
H. Dahrouj and W. Yu, ``Coordinated beamforming for the multi-cell multi-antenna wireless system,'' \emph{IEEE Trans. Wireless Commun.}, vol. 9, no. 5, pp. 1748-1759, May 2010.

\bibitem{ref:Complete}
E. Jorswieck, E. Larsson, and D. Danev, ``Complete characterization of the Pareto boundary for the MISO interference channel,'' \emph{IEEE Trans. Signal Process.}, vol. 56, no. 10, pp. 5292-5296, Oct. 2008.

\bibitem{ref:Optimality}
X. Shang, B. Chen, and H. V. Poor, ``Multiuser MISO interference channels with single-user detection: Optimality of beamforming and the achievable rate region,'' \emph{IEEE Trans. Inf. Theory}, vol. 57, no. 7, pp. 4255-4273, July 2011.

\bibitem{ref:Cooperative}
R. Zhang and S. G. Cui, ``Cooperative interference management with multicell downlink beamforming,''  \emph{IEEE Trans. Signal Process.}, vol. 58, no.10, pp. 5450-5458, Oct. 2010.

\bibitem{ref:Optimal}
J. Qiu, R. Zhang, Z. Q. Luo, and S. G. Cui, ``Optimal distributed beamforming for MISO interference channels,'' \emph{IEEE Trans. Signal Process.}, vol. 59, no. 11, pp. 5638-5643, Nov. 2011.

\bibitem{ref:ICI-Aware}
G. Miao, N. Himayat, Y. Li, and S. Talwar, ``Distributed interference-aware energy-efficient power optimization,'' \emph{IEEE Trans. Wireless Commun.}, vol. 10, no. 4, pp. 1323-1333, Apr. 2011.

\bibitem{ref:MIMOInterf}
C. Jiang and L. J. Cimini, Jr., ``Energy-efficient transmission for MIMO interference channels,'' in \emph{Proc. 2012 IEEE WCNC}, pp. 1-6, Apr. 2012.

\bibitem{ref:EEwoPcir}
S. Verdu, ``Spectral efficiency in the wideband regime,'' \emph{IEEE Trans. Inform. Theory}, vol. 48, no. 6, pp. 1319-1343, June 2002.

\bibitem{ref:ellipsoid}
S. Boyd, \emph{EE364b Lecture Notes}. Stanford, CA: Stanford Univ., available online: \url{http://www.stanford.edu/class/ee364b/lectures/ellipsoid_method_slides.pdf}.

\bibitem{ref:Fractional}
S. Schaible, ``Fractional programming,'' \emph{Zeitschrift fur Operations Research}, vol. 27, no. 1, pp. 39-54, 1983.

\bibitem{ref:Convex}
S. Boyd and L. Vandenberghe, \emph{Convex optimization}. Cambridge, U.K.: Cambridge Univ. Press, 2004.






\end{thebibliography}
\end{document}